\documentclass[twocolumn,showpacs,prl]{revtex4}
\usepackage{graphicx}

\usepackage{bm}
\usepackage{amsmath}
\usepackage{amsfonts}
\newcommand{\ch}{{\cal H}}

\newcommand{\tr}{{\rm Tr}}
\newcommand{\ket}[1]{| #1 \rangle}
\newcommand{\bra}[1]{\langle #1 |}
\newcommand{\braket}[2]{\langle #1 | #2 \rangle}

\begin{document}
\title{Complementarity, distillable secret key, and 
distillable entanglement}

\author{Masato Koashi}
\affiliation{Division of Materials Physics, 
Department of Materials Engineering Science,
Graduate School
of Engineering Science, Osaka University, 
1-3 Machikaneyama, 
Toyonaka, Osaka 560-8531, Japan}
\affiliation{CREST Photonic Quantum Information Project, 4-1-8 Honmachi, Kawaguchi, Saitama 331-0012, Japan}

\begin{abstract} 
We consider controllability of two conjugate observables
$Z$ and $X$ by two parties with classical communication.
The ability is specified by 
two alternative tasks, (i) agreement on $Z$ and (ii) preparation of an
 eigenstate of $X$ with use of an extra communication channel. 
We prove that their
 feasibility is equivalent to that of key distillation if the
 extra channel is quantum, and to that of entanglement distillation 
if it is classical. This clarifies the distinction between two
entanglement measures, distillable key and distillable entanglement.

\end{abstract}

\maketitle

When two remote parties Alice and Bob want to communicate a
message over a public channel without disclosing 
it to a third party Eve, it is sufficient for them to have 
a resource called a (secret) key, which is 
a random number that is shared by Alice and Bob secretly from Eve.
In quantum mechanics, a $(\log d)$-bit key 
is described by a tripartite state 
\begin{eqnarray}
 \tau_{ABE}\equiv d^{-1}\sum_{i=0}^{d-1} \ket{ii}\bra{ii}_{AB}\otimes
\rho_E
\end{eqnarray}
with arbitrary state $\rho_E$,
where we assume Hilbert space $\ch_A\otimes \ch_B \otimes \ch_E$
of systems held by the three parties, with a standard basis 
$\{\ket{i}_A\}$ for $\ch_A$ and $\{\ket{i}_B\}$ for $\ch_B$.
Quantum key distribution (QKD) protocols
enable production of the key through 
communication over a quantum channel and an authenticated public
channel. Strictly speaking, they do not provide the state 
$\tau_{ABE}$ but a state $\rho_{ABE}$ very close to $\tau_{ABE}$.
The imperfection is often measured \cite{Renner-Konig04,HHHO05} 
using the trace norm as
\begin{eqnarray}
 \delta_{\rm key}\equiv \|\rho_{ABE}-\tau_{ABE}\|.
\end{eqnarray}
It is not an easy task to bound $\delta_{\rm key}$ 
against Eve with unconditional power by considering all of her options
in a QKD protocol. Hence we often invoke the fact that
Alice and Bob could have done a different (virtual) protocol 
instead of the actual protocol, 
at least from Eve's point of view.

One of successful approaches \cite{Shor-Preskill00} 
is to take an entanglement distillation 
protocol (EDP) \cite{BDSW96} as the virtual protocol, which tries to produce a
$(\log d)$-ebit
maximally entangled state (MES)
\begin{eqnarray}
 \ket{\phi^{\rm mes}}_{AB}\equiv 
d^{-1/2}\sum_{i=0}^{d-1} \ket{ii}_{AB}.
\label{eq:MES}
\end{eqnarray}
Once its feasibility is proved, the security of the QKD protocol
immediately follows since the task of entanglement distillation
is stronger 
than that of key distillation \cite{DEJMPS96,Lo99}. 
In fact, rather unexpectedly, it was shown \cite{HHHJ05}
that it is often strictly stronger, and distillable entanglement 
$E_D(\rho_{AB})$ is strictly smaller than distillable 
key $K_D(\rho_{AB})$. This implies that the security of a QKD
protocol is not necessarily provable by a reduction to an EDP, and 
distillation of a wider class of states were proposed to
restore the applicability \cite{HHHJ05,HHHO05}.

On the other hand, the first proof of unconditional security by 
Mayers \cite{Mayers96} took a 
quite different approach. He considered a virtual protocol
concerning an observable that is ``conjugate'' to the key.
In contrast to the EDP approach, here neither the real protocol
nor the virtual one alone can prove the security. 
Security follows from 
the fact that Alice and Bob can freely choose between the two 
protocols, which cannot be executed at the same time. 
This complementarity approach has been refined 
\cite{Koashi-Preskill03,Koashi05}
to achieve 
the simplicity comparable to the EDP approach. In addition,
it has a unique practical advantage of low demand on  
the characterization of apparatuses. Recently, this has lead
to the security proof of efficient QKDs using practical sources 
{\em and} detectors \cite{Koashi06,AYKI06}. 

In this paper, 
we first show that this complementarity scenario is not merely 
a tool to prove the security, but it captures exactly what the 
key distillation is, by proving that there exists a corresponding
 complementarity task whenever key can be distilled. 
Then we also show that a slightly different complementarity task,
aimed at the same goal but with the available resource restricted,
is equivalent to entanglement distillation.
These results imply that the distillable key 
$K_D(\rho_{AB})$ and the distillable entanglement $E_D(\rho_{AB})$
have nice alternative definitions in the complementarity scenario,
which clarifies the physical meaning of the difference between the 
two quantities.

We first formulate the complementarity scenario
essentially used in the latest version \cite{Koashi05}
of the security arguments,
which here we call 
{\em complementary control of a $(\log d)$-bit observable}.
We consider a pair of protocols, the primary and the secondary,
between which Alice and Bob can choose to execute.
The two protocols are roughly described as follows. 
In the primary protocol, they communicate over a classical
channel, and then Alice measures a $(\log d)$-bit local observable $Z$,
while Bob tries to guess its outcome. In the secondary protocol,
they perform the same classical communication, but after that
Alice tries to prepare an eigenstate of 
an observable $X$, which is conjugate of $Z$. In doing so,
we allow Bob to help Alice through
 an extra quantum (or classical) channel.


More precisely,   
we require that the choice between the alternative protocols
can be postponed after the end of the classical communication. 
At this point, we assume that 
the standard basis $\{\ket{i}_A\}_{i=0,\ldots,d-1}$ of $\ch_A$ corresponds 
to the observable $Z$. 
If they choose the primary protocol, Alice measures $\ch_A$ 
(system $A$)
on $\{\ket{i}_A\}_{i=0,\ldots,d-1}$ and Bob conducts a local 
operation on his entire systems, resulting in the state of 
systems $AB$
being $\sum_{ij}p_{ij}\ket{ij}\bra{ij}_{AB}$. The error in this 
protocol is given by 
\begin{eqnarray}
 \delta_Z\equiv 1-\sum_i p_{ii}.
\end{eqnarray}
If they choose the secondary protocol, Alice and Bob cooperate over
the extra channel in order to prepare system $A$ in state 
 $\ket{0_X}_A\equiv d^{-1/2}\sum_i \ket{i}_A$. When they end up in 
state $\sigma_A$, we define its error by 
\begin{eqnarray}
 \delta_X\equiv 1-{}_A\bra{0_X}\sigma_A\ket{0_X}_A.
\label{eq:deltax}
\end{eqnarray}
Of course, it would be meaningless if we allowed Alice to 
discard the contents of system $A$ and prepare $\ket{0_X}_A$
from scratch. In order to claim that they really have 
created an eigenstate 
of $X$, conjugate of $Z$,
we require that their operation over the extra channel
should commute with the observable $Z$, namely, it preserves every eigenstate
$\ket{j}_A$. We call it the {\em nondisturbing condition}.

We now show two theorems implying 
that this scenario is essentially equivalent to 
key distillation, as depicted in Fig.~1. In the proofs, 
we use the fidelity \cite{Jozsa94}
$F(\rho,\sigma)\equiv \|\sqrt{\rho}\sqrt{\sigma}\|^2$
as well as the trace distance. Both measures are monotone
under quantum operations, and they are related
by $2(1-\sqrt{F})\le \|\rho-\sigma\|\le 2\sqrt{1-F}$ \cite{Fuchs-Graaf97}.
The fidelity is useful because of the existence of 
extensions $\ket{\phi_\rho}$ and $\ket{\phi_\sigma}$ 
satisfying $|\braket{\phi_\rho}{\phi_\sigma}|^2=F$,
whereas the trace distance obeys the triangle inequality.
Eq.~(\ref{eq:deltax}) can be written as 
$F(\sigma_A, \ket{0_X}\bra{0_X}_A)=1-\delta_X$.

\begin{figure}
\center{\includegraphics[width=.95\linewidth]{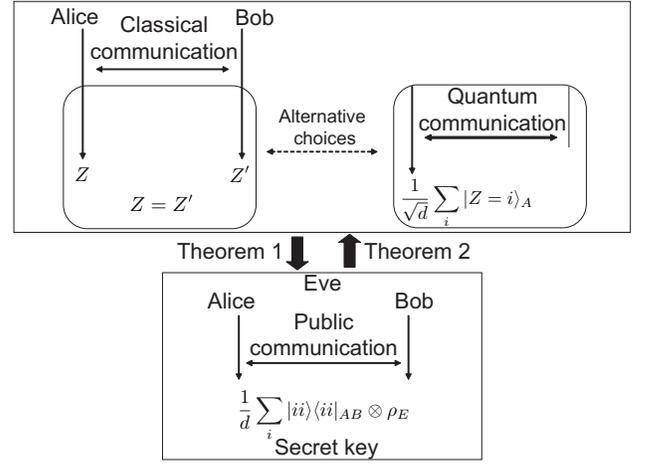}}
\caption{Equivalence between complementary control with an extra
 quantum channel and key distillation.
\label{fig:key}
}
\end{figure} 

The security argument in \cite{Koashi05} 
is essentially given by the following 
theorem.

\noindent
{\bf Theorem 1}. If complementary control of a $(\log d)$-bit observable
with errors $\delta_Z$ and $\delta_X$
is possible with an extra quantum channel, then the primary 
protocol yields a $(\log d)$-bit key with imperfection 
$\delta_{\rm key}\le 2\delta_Z+2\sqrt{\delta_X}$.

Proof. Suppose that Alice and Bob run the primary or the secondary protocol
in the presence of Eve, leading to  
the final states 
$\rho_{ABE}=\sum_{ij}p_{ij}\ket{ij}\bra{ij}_{AB}\otimes\rho_E^{(ij)}$ 
or $\sigma_{AE}$, respectively. 
Suppose that, after the secondary protocol,
(a) we measure system $A$ on the basis $\{\ket{i}_A\}$ to obtain 
$\sigma'_{AE}=\sum_{i}q_{i}\ket{i}\bra{i}_{A}\otimes\rho_E^{(i)}$,
and then (b) copy the outcome onto system $B$, resulting in 
$\sigma'_{ABE}=\sum_{i}q_{i}\ket{ii}\bra{ii}_{AB}\otimes\rho_E^{(i)}$.
We will show that $\rho_{ABE}$ and $\tau_{ABE}$ are both close to 
$\sigma'_{ABE}$. 

Since $\sigma_A\equiv \tr_E\sigma_{AE}$ satisfies Eq.~(\ref{eq:deltax}),
there exists a state $\tau_{AE}\equiv
\ket{0_X}\bra{0_X}_A\otimes
\rho_E$ with $F(\sigma_{AE},\tau_{AE})=1-\delta_X$, and hence
$\|\sigma_{AE}-\tau_{AE}\|\le 2\sqrt{\delta_X}$.
If we apply the steps (a) and (b) to state $\tau_{AE}$, 
the final state is an ideal key $\tau_{ABE}=d^{-1}
\sum_{i}\ket{ii}\bra{ii}_{AB}\otimes\rho_E$. Therefore,
$\|\sigma'_{ABE}-\tau_{ABE}\|\le 2\sqrt{\delta_X}$.

Thanks to the nondisturbing condition, 
$\sigma'_{AE}=\tr_B(\rho_{ABE})$, and hence 
$\sigma'_{ABE}=\sum_{ij}p_{ij}\ket{ii}\bra{ii}_{AB}\otimes\rho_E^{(ij)}$.
Then, direct calculation leads to
$\|\sigma'_{ABE}-\rho_{ABE}\|=2\delta_Z$,
proving Theorem 1.

Next, we show that the opposite direction is also true if there is 
no restriction to Eve's power. Here, we assume the following for 
Eve with no restriction. 
Let us represent the entire data transmitted over the public
 communication by variable $\omega$. We assume that Alice, Bob, and 
Eve each has the record of $\omega$, and hence $\ch_E$ is decomposed 
as $\ch_E=\oplus_\omega \ch_E^{(\omega)}$.
In principle, by using large auxiliary systems $A'$ and $B'$,
Alice and Bob can do the same 
key distillation coherently without discarding any subsystems.
We assume that Eve can collect everything that is not possessed by
Alice and Bob. This ensures that the final state for a particular value
 of $\omega$ is a pure state $\ket{\Phi_\rho^{(\omega)}}_{ABEA'B'}$,
and the overall state is $\rho_{ABEA'B'}=\oplus_\omega p_\omega
\ket{\Phi_\rho^{(\omega)}}\bra{\Phi_\rho^{(\omega)}}$.
Tracing out systems $A'B'$ gives state 
$\rho_{ABE}=\oplus_\omega p_\omega  \rho_{ABE}^{(\omega)}$.

Now we can prove the following theorem.

\noindent
{\bf Theorem 2}. If a $(\log d)$-bit key with imperfection 
$\delta_{\rm key}$ can be distilled against Eve with no restriction,
then complementary control of a $(\log d)$-bit observable
with an extra quantum channel is possible
with errors $\delta_Z\le \delta_{\rm key}/2$ and 
$\delta_X\le \delta_{\rm key}-(\delta_{\rm key}/2)^2$.

Proof. We regard the key distillation protocol as the primary 
protocol. Then $\delta_Z\le \delta_{\rm key}/2$ is trivial.
Before stating the secondary protocol, we need
the following observations. 
In the assumption $\|\rho_{ABE}-\tau_{ABE}\|=\delta_{\rm key}$,
$\tau_{ABE}$ may not be a direct sum over $\omega$. But we can
define such a state  
$\tau'_{ABE}=\oplus_\omega p'_\omega  \tau_{ABE}^{(\omega)}$ by
applying decoherence to $\tau_{ABE}$. Since the same decoherence
operation does not alter $\rho_{ABE}$, we have 
$\|\rho_{ABE}-\tau'_{ABE}\|\le \delta_{\rm key}$,
or $F(\rho_{ABE},\tau'_{ABE})\ge (1-\delta_{\rm key}/2)^2$.
Then, there exists an extension of $\tau'_{ABE}$ taking the 
form of $\tau'_{ABEA'B'}= \oplus_\omega p'_\omega
\ket{\Phi_\tau^{(\omega)}}\bra{\Phi_\tau^{(\omega)}}_{ABEA'B'}$,
satisfying $F(\rho_{ABEA'B'},\tau'_{ABEA'B'})\ge (1-\delta_{\rm key}/2)^2$.

Since $\tau_{ABE}^{(\omega)}=d^{-1}\sum_i 
\ket{ii}\bra{ii}_{AB}\otimes\rho_E^{(\omega)}$, state 
$\ket{\Phi_\tau^{(\omega)}}$ must be written 
in the form of 
$d^{-1/2}\sum_i \ket{i}_{A} \ket{\phi^{(\omega)}_i}_{A'BB'E}$
with $\tr_{A'BB'}\ket{\phi^{(\omega)}_i}\bra{\phi^{(\omega)}_i}
=\rho_E^{(\omega)}$, which is independent of $i$.
This implies the existence of unitaries $\{U_{A'BB'}^{(\omega,i)}\}$
satisfying $U_{A'BB'}^{(\omega,i)}\ket{\phi^{(\omega)}_i}_{A'BB'E}
=\ket{\phi^{(\omega)}_0}_{A'BB'E}$. 
If we define $U_{AA'BB'}^{(\omega)}\equiv \sum_i \ket{i}\bra{i}_A
\otimes U_{A'BB'}^{(\omega,i)}$, we see 
$$
U_{AA'BB'}^{(\omega)}\ket{\Phi_\tau^{(\omega)}}
=d^{-1/2}\sum_i \ket{i}_{A} \ket{\phi^{(\omega)}_0}_{A'BB'E},
$$
where the state of system $A$ is $\ket{0_X}_A$.

Hence we can construct the secondary protocol as follows:
After the coherent version of the key distillation protocol,
using the record of $\omega$,
Alice and Bob apply $U_{AA'BB'}^{(\omega)}$ using an extra 
quantum channel. The form of $U_{AA'BB'}^{(\omega)}$ obviously 
satisfies the nondisturbing condition.
If the state after the key distillation protocol
was $\tau'_{ABEA'B'}$, the protocol would produce $\ket{0_X}_A$ exactly.
Thus, for state $\rho_{ABEA'B'}$, the output $\sigma_A$ should 
satisfy ${}_A\bra{0_X}\sigma_A\ket{0_X}_A\ge (1-\delta_{\rm key}/2)^2$,
namely, $\delta_X\le \delta_{\rm key}-(\delta_{\rm key}/2)^2$.

The two theorems indicate that the complementarity scenario is 
a powerful tool for QKD, namely, there 
is no fundamental limitation in applying the scenario to 
prove the security of QKD protocols.
They also show that distillable key $K_D(\rho_{AB})$ \cite{HHHO05}
of a bipartite state $\rho_{AB}$ 
can be also defined in the complementarity scenario.
Let us introduce the asymptotic yield of complementary control 
$Y_Q$, where the subscript signifies that the 
extra channel is quantum. We define $Y_Q(\rho_{AB})$ to 
be the supremum of real numbers $y$ with which the following 
statement holds true. Starting with $\rho_{AB}^{\otimes
n}$, complementary control of a 
$(\log d_n)$-bit observable is possible with errors 
$(\delta^{(n)}_Z,\delta_X^{(n)})$, where
$\delta^{(n)}_Z \to 0$, $\delta^{(n)}_X \to 0$, and
$\log d_n/n \to y$ for 
$n\to \infty$. With this definition, Theorem 1 implies 
$K_D(\rho_{AB})\ge Y_Q(\rho_{AB})$ while Theorem 2 implies
$K_D(\rho_{AB})\le Y_Q(\rho_{AB})$, leading to
\begin{eqnarray}
 K_D(\rho_{AB})= Y_Q(\rho_{AB}).
\label{eq:KDandYQ}
\end{eqnarray}

\begin{figure}
\center{\includegraphics[width=.95\linewidth]{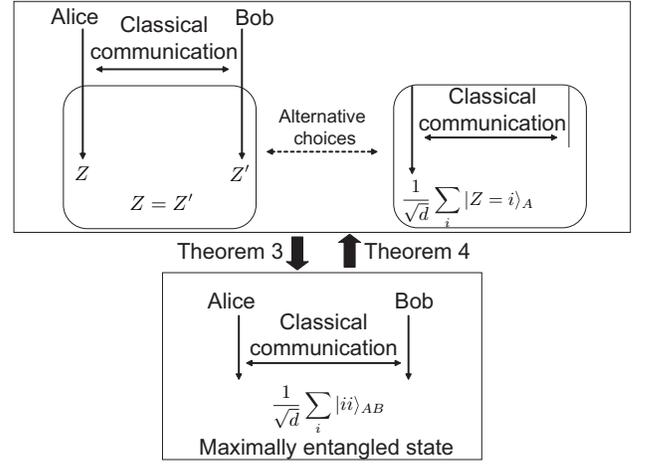}}
\caption{Equivalence between complementary control with an extra
 classical channel and entanglement distillation.
\label{fig:2}
}
\end{figure} 

Next, let us consider a slightly different scenario, in which 
Alice and Bob are allowed to communicate only classically
in the secondary protocol. Then we can find a close connection 
to distillation of the $(\log d)$-ebit maximally entangled state 
defined in Eq.~(\ref{eq:MES}), 
as depicted in Fig.~\ref{fig:2}. For output state $\rho_{AB}$,
we measure the imperfection in the distillation task by 
\begin{eqnarray}
 \delta_{\rm ent}\equiv \|\rho_{AB}-\tau^{\rm ent}_{AB}\|
\end{eqnarray}
with $\tau^{\rm ent}_{AB}\equiv \ket{\phi^{\rm mes}}\bra{\phi^{\rm mes}}$.
Then we can prove the following.

\noindent
{\bf Theorem 3}. If complementary control of a $(\log d)$-bit observable
with errors $\delta_Z$ and $\delta_X$
is possible with an extra classical channel, then 
it is possible to distill a $(\log d)$-ebit maximally entangled
state with imperfection $\delta_{\rm ent}\le
4\sqrt{\delta_Z(1-\delta_Z)}
+2\sqrt{\delta_X}$. 

Proof. In the primary protocol, Alice and Bob's 
operation after the classical
communication can be coherently done by enlarging the size of 
systems $A'B'$, namely, it can be done by a unitary operation 
$V= V_{AA'}\otimes V_{BB'}$. 
Let $\sigma_{AA'BB'}$ be the state after the application 
of $V$. The error $\delta_Z$ implies that
\begin{eqnarray}
 \tr[Q_{AB}^{\rm cor}\sigma_{AA'BB'}]=1-\delta_Z
\label{eq:smallZ}
\end{eqnarray}
where $Q_{AB}^{\rm cor}\equiv \sum_i \ket{ii}\bra{ii}_{AB}$ 
is the projection onto the subspace with no errors.

From the state $\sigma_{AA'BB'}$, Alice and Bob can 
undo the unitary by applying $V^{-1}$, 
going back to the state just after the
classical communication. From here, Alice and Bob can choose 
to conduct the secondary protocol using the extra classical 
channel to produce state $\rho_A=\Lambda(\sigma_{AA'BB'})$,
 where we write the whole quantum operation starting from 
$V^{-1}$ by a CPTP map $\Lambda$. The error $\delta_X$
implies that
\begin{eqnarray}
\|\Lambda(\sigma_{AA'BB'})-\ket{0_X}\bra{0_X}_A\|\le 2\sqrt{\delta_X}.
\label{eq:smallX}
\end{eqnarray}

We construct a distillation protocol as follows.
Alice and Bob conduct the coherent version of 
the primary protocol, resulting in  $\sigma_{AA'BB'}$.
Bob further uses an auxiliary system $C$ with dimension $d$,
prepared in state $\sigma_C\equiv\ket{0}\bra{0}_C$. 
He copies the contents of system $B$ onto system $C$
by unitary $U_{BC}^{\rm cpy}: \ket{j0}_{BC}\mapsto \ket{jj}_{BC}$,
resulting in state
\begin{eqnarray}
 \sigma'_{AA'BB'C}=U_{BC}^{\rm cpy}(\sigma_{AA'BB'}\otimes \sigma_C)
U_{BC}^{\rm cpy\dagger}.
\end{eqnarray}
Alice and Bob then apply $\Lambda$ on systems $AA'BB'$ 
using the extra classical channel
to achieve the final state 
$\rho'_{AC}=\Lambda(\sigma'_{AA'BB'C})$.

We now prove that $\rho'_{AC}$ is close to $\tau_{AC}^{\rm ent}$.
Consider the state defined by
 \begin{eqnarray}
 \sigma''_{AA'BB'C}=U_{AC}^{\rm cpy}(\sigma_{AA'BB'}\otimes \sigma_C)
U_{AC}^{\rm cpy\dagger}.
\end{eqnarray}
with $U_{AC}^{\rm cpy}: \ket{j0}_{AC}\mapsto \ket{jj}_{AC}$.
Using Eq.~(\ref{eq:smallZ}) and the obvious relation 
$U_{AC}^{\rm cpy}Q_{AB}^{\rm cor}=U_{BC}^{\rm cpy}Q_{AB}^{\rm cor}$,
we can show \cite{note070406} that 
$\|\sigma'_{AA'BB'C}-\sigma''_{AA'BB'C})\|\le 
4\sqrt{\delta_Z(1-\delta_Z)}$
and hence
\begin{eqnarray}
 \|\rho'_{AC}-\Lambda(\sigma''_{AA'BB'C})\|\le
  4\sqrt{\delta_Z(1-\delta_Z)}.
\label{eq:boundZ}
\end{eqnarray}
On the other hand, the nondisturbing condition implies that 
there is no difference whether we apply 
$U_{AC}^{\rm cpy}$ before or after the application of $\Lambda$.
This leads to 
\begin{eqnarray}
 \Lambda(\sigma''_{AA'BB'C})=
U_{AC}^{\rm cpy}(\Lambda(\sigma_{AA'BB'})\otimes \sigma_C)
U_{AC}^{\rm cpy\dagger}.
\end{eqnarray}
Then, using Eq.~(\ref{eq:smallX}), we have
\begin{eqnarray}
 \|\tau_{AC}^{\rm ent}-\Lambda(\sigma''_{AA'BB'C})\|\le 2\sqrt{\delta_X}.
\end{eqnarray}
Combined with Eq.~(\ref{eq:boundZ}), it proves Theorem 3.

The opposite direction is trivial, and it is stated as follows
(proof omitted).

\noindent
{\bf Theorem 4}.
If a $(\log d)$-ebit maximally entangled
state with imperfection $\delta_{\rm ent}$ 
can be distilled,
then complementary control of a $(\log d)$-bit observable
with an extra classical channel is possible
with errors $\delta_Z\le \delta_{\rm ent}/2$ and 
$\delta_X\le \delta_{\rm ent}-(\delta_{\rm ent}/2)^2$.

If we define the asymptotic yield $Y_C(\rho_{AB})$ with 
an extra classical channel as we defined $Y_Q$
before, Theorems 3 and 4 lead to  
\begin{eqnarray}
 E_D(\rho_{AB})= Y_C(\rho_{AB}),
\end{eqnarray}
which shows that the distillable entanglement $E_D$ also
has an alternative definition in the complementarity scenario.
Together with Eq.~(\ref{eq:KDandYQ}), now we see that 
distillable key and distillable entanglement can be regarded
as achievable yields of the same task, carried out under 
different conditions. This gives a clear distinction between 
the two entanglement measures.
Both are related to the potential to carry out 
two mutually exclusive tasks concerning a pair of conjugate 
observables $Z$ and $X$, using the same classical communication.
One task is to share the value of $Z$, and the other one 
is to drive the state into an eigenstate of $X$. The latter 
task naturally requires additional communication, and this 
is where the difference between the two quantities shows up. 
If we insist that it also must be classical and hence both
tasks are feasible with only classical communication, 
the achievable size of the observables tallies with the
distillable entanglement. If we place no such requirement, 
then the achievable size matches the distillable key.
This may be understandable because if the key is 
actually distilled, the task for $X$ is never carried out 
and hence there is no concern about 
what resources are required to
carry it out.

We have seen that the complementarity scenario can explain 
two of the few operationally-defined entanglement measures,
which shows its significance in understanding quantum
entanglement. It is interesting to see whether we can 
also define yet another operationally-defined measure,
entanglement cost \cite{HHT01} in a complementarity scenario. 
The task of the complimentary control defined here is 
merely one of many possible ways to quantify abilities  
related to the concept of complementarity, and 
it is worth seeking other tasks, for example, the ones 
retaining the symmetry between two conjugate observables.
 
The author thanks N.~Imoto and T.~Yamamoto for helpful 
discussions. This work was supported by a MEXT Grant-in-Aid 
for Young Scientists (B) 17740265.


\bibliographystyle{apsrev}

\end{document}